# Phase evolution and superconductivity enhancement in Se-substituted MoTe$_2$ thin films

Peiling Li[1,2,†], Jian Cui[1,†], Jiadong Zhou[3,†,*], Dong Guo[1], Zhenzheng Zhao[1], Jian Yi[4], Jie Fan[1,5], Zhongqing Ji[1,5], Xiunian Jing[1,6], Fanming Qu[1], Changli Yang[1,6], Li Lu[1,6], Junhao Lin[7,8,*], Zheng Liu[3,*], and Guangtong Liu[1,5*]

[1]Beijing National Laboratory of Condensed Matter Physics, Institute of Physics, Chinese Academy of Sciences, Beijing 100190, China

[2]University of Chinese Academy of Sciences, Beijing 100049, China

[3]School of Materials Science and Engineering, Nanyang Technological University, Singapore 639798, Singapore

[4]Ningbo Institute of Industrial Technology, Chinese Academy of Sciences, Ningbo 315201, China

[5]Songshan Lake Materials Laboratory, Dongguan, Guangdong 523808, China

[6]Collaborative Innovation Center of Quantum Matter, Beijing 100871, China

[7]Department of Physics, Southern University of Science and Technology, Shenzhen 518055, China

[8]Shenzhen Key Laboratory of Quantum Science and Engineering, Shenzhen 518055, China

†These authors contributed equally to this work. Correspondence and requests for materials should be addressed to J.Z (jzhou012@e.ntu.edu.sg), J.L (email: lin.junhao.stem@gmail.com), Z.L. (email: z.liu@ntu.edu.sg) and G.L. (email: gtliu@iphy.ac.cn)

**Abstract**

The strong spin-orbit coupling (SOC) and numerous crystal phases in few-layer transition metal dichalcogenides (TMDCs) $MX_2$ (M=W, Mo, and X=Te, Se, S) has led to a variety of novel physics, such as Ising superconductivity and quantum spin Hall effect realized in monolayer $2H$- and $T_d$-$MX_2$, respectively. Consecutive tailoring of the $MX_2$ structure from $2H$ to $T_d$ phase may realize the long-sought topological superconductivity in one material system by incorporating superconductivity and quantum spin Hall effect together. In this work, by combing Raman spectrum, X-ray photoelectron spectrum (XPS), scanning transmission electron microscopy imaging (STEM) as well as electrical transport measurements, we demonstrate that a consecutively structural phase transitions from $T_d$ to $1T$' to $2H$ polytype can be realized as the Se-substitution concentration increases. More importantly, the Se-substitution has been found to notably enhance the superconductivity of the $MoTe_2$ thin film, which is interpreted as the introduction of the two-band superconductivity. The chemical constituent induced phase transition offers a new strategy to study the $s_{+-}$ superconductivity and the possible topological superconductivity as well as to develop phase-sensitive devices based on $MX_2$ materials.

**Introduction**

Two-dimensional (2D) transition metal dichalcogenides (TMDCs) have attracted extensive interests in the fields of condensed matter physics and material sciences in recent years[1,2]. These materials having the chemical formula $MX_2$ (M =W, Mo and X=Te, Se, S) can crystallize in a variety of polytypic structures such as $2H$, $1T$, $1T'$, and $T_d$ phases. The strong spin-orbit coupling and diverse structures in few-layer $MX_2$ has led to the discovery of fundamentally new physics, including Ising superconductivity[3,4] in gated-$MoS_2$ and monolayer $NbSe_2$ with $2H$ structure; quantum spin Hall effect[5-8] in monolayer $WTe_2$ with $1T'$ phase; nonlinear Hall effect in few-layer $T_d$-$WTe_2$[9-10]; as well as asymmetric spin-orbit coupling in few-layer superconducting $T_d$-$MoTe_2$[11]. On the other hand, phase engineering could bring exciting new applications for $MX_2$, such as electronic, optoelectronic and valleytronic sensors[2,12-14]. Therefore, it is highly desired to develop a controllable technique to tailor a specific phase in $MX_2$, which may open up new possibilities for fundamental studies[5,15,16] and phase-sensitive devices applications[17,18] based on $MX_2$ thin films.

To date, much effort has been devoted to realizing the $2H$-to-$1T'$ phase transition in $MX_2$. The methods involve heating[19,20], element doping[21,22], chemical modifications[23,24], laser irradiation[25], strain[26], and electrostatic gating[27]. Recently, the $1T'$-to-$T_d$ phase transition driven by pressure[28], thermo[29,30], as well as dimensionality[11,31] in $MX_2$ has been reported. However, these strategies can only realize $2H$-to-$1T'$ or $1T'$-to-$T_d$ phase transition. On the other hand, thank to the development of chemical vapor deposition (CVD) on the controlled synthesis of atomically thin $MX_2$ with different phases[32], a complete phase tuning from 2H to $T_d$ becomes promising.

In this paper, we demonstrate a controlled strategy to fabricate specific phase in Se-substituted $MoTe_2$ thin films. We show that a series of $MoSe_xTe_{2-x}$ (with $0 \leqslant x \leqslant 1.3$) samples with fully tunable chemical compositions and electrical properties can be synthesized by a facile CVD method. The structural characterizations including Raman spectrum, STEM, XPS and electrical transport measurements demonstrate that a transition from $T_d$ to $1T'$ phase occurs around x~0.8, and $1T'$ to $2H$ phase around x~1.1. More importantly, we find that the Se substitution has a significant enhancement effect on the superconductivity. With the Se content increased from x=0 to x=1.0, the superconducting transition temperature ($T_c$) shifts from 2.1 K to 3.6 K; and for x >1.0, the superconductivity is completely suppressed due to the structural phase transition from the metallic $1T'$ phase to the semiconducting $2H$ polytype.

**Results**

**Growth of $MoSe_xTe_{2-x}$ thin films.** Se-substituted $MoTe_2$ samples with nominal composition $MoSe_xTe_{2-x}$ (0≤x≤1.5) are synthesized by the molten-salt assisted chemical vapor deposition (CVD) method[32]. More information about the sample growth is detailed in the Methods section and Supplementary Figure 1. Figures 1a – 1c show the representative optical images of the as-synthesized $MoSe_xTe_{2-x}$ films with different thickness and shapes. The thickness of the $MoSe_xTe_{2-x}$ can be controlled via the growth time to reach a monolayer limit. Notably, when the ratio between Se and Te is larger than 1 (x>1), the dominant shape of the as-synthesized sample is triangle or hexagonal, suggesting the $2H$ crystal structure, while rectangular shape is always observed when the ratio of Se/Te is less than 1 (x<1). Furthermore, Raman spectroscopy is used to study the phase transition between the $T_d/1T'$ phase and the $2H$ phase. Figure 1d shows the Raman spectra of as-synthesized $MoSe_xTe_{2-x}$ samples with different Se contents (x) as determined by X-ray photoelectron spectroscopy (XPS). In the pristine $MoTe_2$ (x=0), Raman peaks are observed at 95, 113, 131 and 163 cm$^{-1}$, which are the same as the previous reports on the $A_1$ and $A_2$ modes in $T_d$-$MoTe_2$[33]. After Se substitution, two main peaks of $T_d$-$MoTe_2$ at 113 and 131 cm$^{-1}$ exhibit clear red-shift induced by the introduction of Se atom. From x=0.3 to x=0.8, all peaks show blue-shift relative to pristine $T_d$-MoTe$_2$. Combining the following scanning transmission electron microscopy (STEM) data, we find that the phase becomes the 1$T'$ phase at x=0.8. Therefore, the blue-shift is ascribed to the phase transition from the $T_d$ to the 1$T'$ phase. As for x≥0.9, two new peaks appear at 122 cm$^{-1}$ and 178 cm$^{-1}$ related to $E_{1g}$ and $A_{1g}$ mode, indicating the appearance of 2$H$ phase. Furthermore, the blue-shift guide line of peak $^2A_2$ extends to x=1.1, indicating the formation of 1$T'$-2$H$ mixed phase for 0.9≤x≤1.1.

**Structural characterization.** Aberration corrected annular dark-field scanning transmission electron microscopy imaging (ADF-STEM) is applied to characterize the phase evolution of the as-grown MoSe$_x$Te$_{2-x}$. We investigate representative MoSe$_x$Te$_{2-x}$ monolayers with triangular and rectangular shapes from Se-rich and Te-rich growth conditions, respectively. Figure 2a shows the atomically resolved ADF-STEM image of a local region of the triangular monolayer grown in a Se-rich condition. The image shows hexagonal lattice pattern with discrete intensity distribution, similar to other ternary monolayer alloy in the 2$H$ phase due to the prismatic trigonal lattice structure[32], where the anion sites have different configurations between Se and Te atom resulting in various intensity. The hexagonal lattice feature of the 2$H$ phase is further revealed by the fast Fourier transformation (FFT) pattern (inset of Fig. 2a). To quantitatively analyze the chemical stoichiometry of the monolayer, we statistically map out the intensity of each atomic column since the image intensity is directly related to the atomic weight of the imaged specie[34]. Figure 2b shows the intensity histogram of all atomic columns in Fig. 2a. The intensity distribution of the cation site converges to a single peak, which is identified as Mo element. On the other hand, the intensity distribution of the anion sites splits into four peaks, which are assigned as single Se, Se$_2$ column, Se + Te column, and Te$_2$ columns in the sequence of their atomic weight. For instance, statistically the Mo intensity is slightly lower than the $Se_2$ column as reported previously in $MoSe_2$ monolayer[35] due to its atomic weight, therefore the brighter intensity in the anion sites beyond the $Se_2$ column can only be attributed to the presence of heavier Te atoms, resulting in the assignments of the last two peaks in the anion intensity histogram as Se + Te and $Te_2$ columns. Note that single Te may also present in the anion site. However, its contribution is quite small which can not be resolved statistically in the intensity histogram, thus negligible in our local chemical stoichiometry analysis. The local chemical stoichiometry of the region shown in Fig. 2a is estimated to be $MoSe_{1.52}Te_{0.48}$ through atom-by-atom counting, consistent with the energy dispersive spectroscopy (EDS) result on a larger region of the same sample as shown in Fig. 2d, which shows $MoSe_{1.4}Te_{0.6}$. Both local atomic counting and EDS measurement unambiguously demonstrate the $2H$ phase is dominant if Te is heavily substituted by Se in $MoTe_2$.

In contrast, we investigate the $MoSe_xTe_{2-x}$ flake with a rectangular shape grown in a Te-rich condition. Figure 2c shows the atomic resolution ADF-STEM image of a local region of the monolayer. In sharp contrast to the $2H$ phase observed in Se-rich grown sample, the lattice shows a quasi-one-dimensional chain-like structure, where pairs of Mo atom pull together towards each other forming a distorted $1T'$ phase, a typical lattice feature of the $1T'/T_d$ phase (the monolayer of the $1T'$ and the $T_d$ phase is essentially the same)[36], as further revealed by the FFT pattern (inset of Fig. 2c). The chemical stoichiometry of the flake is quantified by the EDS to be $MoSe_{0.8}Te_{1.2}$, indicating a Se:Te ratio smaller than 1. Combined with the atomic resolution images and the EDS data from the $2H$ and the $1T'$ phases, we confirm the phase transition from the $2H$ to $1T'$ phase occurs as the ratio of Se:Te decreases to 0.8 (x~0.9) in $MoSe_xTe_{2-x}$ samples, consistent with the above Raman and XPS data shown below.

To assess the precise stoichiometry and the phase information, the as-synthesized $MoSe_xTe_{2-x}$ are further analyzed via X-ray photoelectron spectroscopy (XPS). Figure 3a shows the typical XPS result of MoSeTe on the $SiO_2$/Si wafer. Five elements including Mo, Se, Te, C, and O can be clearly discerned in the XPS spectra. The signals of C and O are assigned to the charge reference and the $SiO_2$ substrate, respectively. The elements of Mo, Se and Te are regarded to originate from Se-substituted $MoTe_2$, and the atomic ratio of Mo:Se:Te is quantified to be 1:1:1 through the computation of peak areas. To extract the phase information of $MoSe_xTe_{2-x}$, we systematically investigate the XPS spectra of Mo, Se, and Te as a function of x. The corresponding XPS spectra are shown in Figs. 3b~3d. Each set of the XPS spectra are analyzed by deconvolution according to the peaks of Mo, Se, and Te, respectively. The doublet peaks located around 227.7 and 231.0 eV in Fig. 3b are observed for x=0.3, which arise from the Mo $3d_{5/2}$ and Mo $3d_{3/2}$ orbitals of the $T_d$-$MoTe_2$, consistent with previous reports[37]. With x further increased, another pair of doublets at 228.5 and 231.5 eV denoted by the dark purple color begin to emerge, which is attributed to Mo $3d_{5/2}$ and Mo $3d_{3/2}$ within the $1T'$ phase. Obviously, the phase ratio of $1T'$/$T_d$ gets larger with increasing Se concentrations. At x=0.8, the $1T'$ phase is nearly comparable to the $T_d$ phase. Then the $1T'$ phase becomes the dominant phase as x≥0.9, indicating a phase transition from the $T_d$ to the $1T'$ phase occurs around x=0.8. With x≥1.2, the doublet peaks at 229.0 and 232.1 eV belonging to Mo $3d_{5/2}$ and Mo $3d_{3/2}$ in $2H$-$MoTe_2$, become conspicuous. The observed phase evolution from $T_d$ to $1T'$ to $2H$ phase in $MoSe_xTe_{2-x}$ is further supported by the Te $3d_{5/2}$ and Se 3d XPS spectra as shown in Figs. 3c and 3d, respectively. From Fig. 3c, one can clearly see that the dominant phase gradually changes from the $T_d$ phase (featured by the Te $3d_{5/2}$ peak at 572.1 eV) to the $1T'$ phase (peak at 572.7 eV) around x=0.8, and finally changes into the $2H$ phase (peak at 573.2 eV) with x≥1.2. Similar phase transition can also be observed in Te $3d_{3/2}$ XPS spectra, as shown in Supplementary Figure 2. From the Se 3d XPS spectra shown in Fig. 3d, one can see that the 2*H* phase emerges at x~0.7 and becomes dominant phase at x~1.1. In addition to Mo, Se, and Te, the signals of $MoO_2$, $MoO_3$, $SeO_2$, and $TeO_2$ are found to exist in the XPS spectra, which is due to the sample oxidation during the device fabrication and transport measurements as previously reported in $MoS_2$[38,39]. Consequently, we can conclude that the phase evolution from $T_d$ to 1*T'* to 2*H* phase in $MoSe_xTe_{2-x}$ can be realized sequentially by tuning the Se:Te ratio.

**Transport properties of few-layer $MoSe_xTe_{2-x}$.** It is known that $T_d/1T'$-$MoTe_2$ is a metal, while 2*H*-$MoSe_2$ is a semiconductor having a direct band gap of ~1.55 eV[40]. Therefore, we can employ the electrical transport measurements to demonstrate the phase evolution from $T_d/1T'$ to 2*H* phase in Se-substituted $MoTe_2$ as a function of Se content x. Figure 4a shows the temperature dependence of the sheet resistance ($R_s$) for $MoSe_xTe_{2-x}$ thin films. The inspection of this figure clearly shows that doping the Te site with Se significantly changes the conducting behavior. At low Se-doping level (x≤1.0), $MoSe_xTe_{2-x}$ indeed displays a metallic behavior (d$R$/d$T$>0) at high temperatures before entering the superconducting state. The residual resistance ratio, RRR = $R_{300K}/R_n$ with $R_{300K}$ as the room-temperature resistance and $R_n$ as the normal-state resistance right above the superconducting transition, varies from 1.43 for x=0 to 1.96 for x=1.0 (Table 1). Furthermore, no appreciable thermal hysteresis was observed in the temperature-dependent sheet resistance $R_s(T)$ curves (Supplementary Figure 3) in superconducting $MoSe_xTe_{2-x}$ samples with x≤1.0, signifying no phase transition from 1*T'* to $T_d$ phase occurs[11,31]. Combined with the above STEM and XPS results, we can conclude that the $T_d$ phase (for x≤0.8) and 1*T'* phase (for x≤1.1) persists to low temperatures in $MoSe_xTe_{2-x}$ thin film. However, with x increased to 1.1 or above, the derivation of $R_s(T)$ curves show a negative sign ($dR/dT<0$) at $T<100$ K, which can be described by the Mott variable-range hopping (VRH) model[41]:

$$R_s(T) = R_0 \exp\left[\left(\frac{T_0}{T}\right)^{1/(d+1)}\right] \quad (1)$$

where $d$ is the dimensionality of a system and $T_0$ is a characteristic temperature. As shown in Supplementary Figure 4, the best fitting leads to $d=2$, indicating two-dimensional (2D) Mott VRH transport in disordered MoSe$_x$Te$_{2-x}$ semiconductors. Furthermore, when all the Te atoms are replaced by Se (x=2), as displayed in Fig. 4a (right axis), $R_s(T)$ increases with decreasing temperature, exhibiting a typical thermally-activated semiconducting behavior. Additionally, the $1T'$-to-$2H$ phase transition is also supported by calculating electron density $n_e$ from the Hall effect measurements. As shown in Supplementary Figure 5, $n_e$ decreases monotonically with increasing Se content, dropping two orders of magnitude from $6\times10^{21}$ cm$^{-2}$ to $2\times10^{19}$ cm$^{-2}$ as x increases from 0 to 1.3. Correspondingly, the room-temperature sheet resistance $R_s(300K)$ increases almost 5 orders of magnitude from 63 Ω to 2.6 MΩ (Table I). Therefore, the phase evolution from $T_d$ to $1T'$ to $2H$ phase can be established in Se-substituted MoTe$_2$ thin films based on the Raman, XPS, STEM, and transport measurements.

More importantly, we find that the superconductivity can be substantially enhanced when partial Te is substituted by Se in MoTe$_2$. Figure 4b plots the normalized sheet resistance $R_s/R_{4K}$ as a function of temperature for pristine MoTe$_2$ and MoSe$_x$Te$_{2-x}$ thin films with x up to 1.0. The superconducting transition temperature $T_c$, defined as the temperature at the midpoint of the resistive transition, gradually rises to 3.6 K as x increases to 1.0. Compared with the pristine bulk MoTe$_2$, nearly 36-fold in $T_c$ is improved in MoSeTe thin films. Figure 4c summaries the phase evolution by displaying $R_s$ as functions of temperature $T$ and Se content x. It is seen that $T_c$ increases monotonically with increasing x, which resembles the case of Te-deficient $T_d$-MoTe$_2$[42].

**Discussion of the enhanced superconductivity**

Superconductivity enhancement has been observed in Te-deficient[42] and S-doped[43,44] $T_d$-MoTe$_2$ bulk samples, where the enhancement mechanism was ascribed to the electron doping effect and the enhanced electron-phonon coupling strength $\lambda_{ep}$, respectively. To access $\lambda_{ep}$ in Se-substituted MoTe$_2$ samples, we firstly calculate the Debye temperature $\Theta_D$ by fitting the normal-state resistivity with the Bloch-Grüneisen function[45]:

$$\rho(T) = \rho(0) + A\left(\frac{T}{\Theta_D}\right)^5 \int_0^{\frac{\Theta_D}{T}} \frac{x^5}{(e^x-1)(1-e^{-x})} dx \qquad (2)$$

where $\rho(0)$ is the residual resistivity and $A$ is a constant. From Fig. 5a, one can see that the experimental data can be described by eq. (2). Figure 5b summarizes the Debye temperature $\Theta_D$ for different Se content. Clearly, $\Theta_D$ depends proportionally on x. Then, $\lambda_{ep}$ can be estimated from the McMillan formula[45]:

$$\lambda_{ep} = \frac{\mu^* \ln\left(\frac{1.45 T_c}{\Theta_D}\right) - 1.04}{1.04 + \ln\left(\frac{1.45 T_c}{\Theta_D}\right)(1 - 0.62\mu^*)} \qquad (3)$$

where $\mu^*$ is the repulsive screened Coulomb potential and is assigned a typical value of 0.1. Figure 5b (right axis) plots the electron-phonon coupling strength $\lambda_{ep}$ for different Se-doping level x. In stark contrast to $\Theta_D$, $\lambda_{ep}$ is insensitive to x and fluctuates around 0.65, suggesting that MoSe$_x$Te$_{2-x}$ are weak-coupling BCS superconductors. Compared to $\lambda_{ep}$=0.31 in bulk MoTe$_2$[43,47],

$\lambda_{ep}$ in Se-substituted MoTe$_2$ thin films has increased more than two-fold, which can qualitatively explain the enhancement of $T_c$. However, it can not explain the Se content dependence of $T_c$.

On the other hand, the enhancement of $T_c$ has been recently observed in pressure-induced two-band superconductivity with the s$^{+-}$ pairing symmetry in bulk MoTe$_2$[48]. Similarly, chemical pressure is introduced in our case when partial larger Te$^{2-}$ (2.21 Å) is substituted by Se$^{2-}$ (1.98 Å)[49]. Therefore, the enhancement of $T_c$ may be explained by calculating pressure-induced larger superconductivity pairing energy[44]. To solidify this assumption, systematic magneto-transport measurements are conducted on pristine $T_d$-MoTe$_2$ and MoSe$_x$Te$_{2-x}$ with x=0.95 thin films under a perpendicular filed $H_\perp$ at different temperatures. Figures 5c and 5d show the temperature dependence of the upper critical field $\mu_0 H_{c2,\perp}(T)$ defined as the field where the resistivity drops to 0.5$R_{ns}$, where $R_{ns}$ is the normal-state sheet resistance. From Fig. 5c, one can see that $\mu_0 H_{c2,\perp}(T)$ of pristine MoTe$_2$ can be well described by the single-band s-wave Werthamer-Helfand-Hohenberg (WHH) model[50] across all temperatures. Compared with the single-band model, MoSe$_{0.95}$Te$_{1.05}$ ($T_c$~3.43 K, Supplementary Figure 6) exhibits a strong enhancement in $\mu_0 H_{c2,\perp}(T)$ at low temperature regime, which can only be fitted by the two-band model with the form[51],

$$a_0[\ln t + U(h)][\ln t + U(\eta h)] + a_1[\ln t + U(h)] + a_2[\ln t + U(\eta h)] = 0$$

$$t = \frac{T}{T_c}, U(x) = \psi\left(\frac{1}{2}+x\right) - \psi\left(\frac{1}{2}\right); \eta = \frac{D_2}{D_1}; h = \frac{H_{c2} D_1}{2\varphi_0 T}; a_0 = \frac{2\varpi}{\lambda_0}; a_1 = 1 + \frac{\lambda_-}{\lambda_0}; a_2 = 1 - \frac{\lambda_-}{\lambda_0};$$

$$\varpi = \lambda_{11}\lambda_{22} - \lambda_{12}\lambda_{21}; \lambda_0 = (\lambda_-^2 + 4\lambda_{12}\lambda_{21})^{\frac{1}{2}}; \lambda_- = \lambda_{11} - \lambda_{22}. \tag{4}$$

where $\psi(x)$ is the digamma function, $\varphi_0 = \frac{h}{2e}$ is the magnetic flux quantum, $D_1$ and $D_2$ represent intraband diffusivities of bands 1 and 2. $\lambda_{11}$ and $\lambda_{22}$ are the intraband couplings between bands

1 and 2. $\lambda_{12}$ and $\lambda_{21}$ describe the interband couplings between bands 1 and 2. If $D_1 = D_2$, eq. (4) reduces to the single-band WHH model. The best fitting gives $D_1=D_2=2.23\times10^{-4}$ m$^2$/s for pristine MoTe$_2$, while $D_1=2.20\times10^{-4}$ and $D_2=3.9\times10^{-5}$ m$^2$/s for MoSe$_{0.95}$Te$_{1.05}$, respectively. Compared with pristine MoTe$_2$, the substitution of Te with Se only introduces a second band with a lower diffusivity $D_2$, but with negligible influence on the original electron diffusion constant $D_1$. The electron diffusion constant $D$ can also be estimated with $D = 1/3 v_f l$ through Hall effect measurements, where $v_f$ and $l$ are Fermi velocity and mean free path. The obtained $D$ are listed in Supplementary Table II. It is found that the value of $D_1$ is consistent with $D$, both of them are of the same order of magnitude ~$10^{-4}$ m$^2$/s, indicating the electrons with the diffusivity $D_1$ are the majority carriers. Therefore, our data suggest that the observed enhancement of $T_c$ is associated with the introduction of low-diffusivity electrons, which may have a larger superconducting gap. We note that much work including theoretical and experimental study is needed to elucidate the mechanism of such superconductivity enhancement.

**Conclusion**

In summary, we have shown an efficient strategy to tailor specific phase and tune the electrical properties of Se-substituted MoTe$_2$ thin film by CVD method. The Raman, XPS, STEM and transport measurements demonstrate that a sequence of phase evolution from $T_d$ to $1T'$ to $2H$ phase can be realized by tuning the Se concentration. More importantly, it is found that Se substitution can notably enhance the superconductivity from $T_c$=0.1 K in bulk pristine MoTe$_2$ to 3.6 K in MoSeTe, which is probably due to the introduction of s$^{+-}$ superconducting order parameter. The Se substitution strategy provides a convenient and powerful way to tailor the structural and physical properties in MX$_2$, paving the way for future phase-sensitive devices applications.

**Materials and Methods**

**CVD synthesis of highly quality Se-substituted $MoSe_xTe_{2-x}$ films.** The $MoSe_xTe_{2-x}$ samples were synthesized via molten-salt assisted CVD method inside a furnace with a 1 inch diameter quartz tube. Specifically, one alumina boat containing precursor powder ($NaCl:MoO_3$=1:5) was put in the center of the tube. The Si substrate with a 285 nm thick $SiO_2$ on top was placed on the alumina boat with polished side faced down. Another alumina boat containing Te and Se powder with different ratio was put on the upstream side of quartz tube at a temperature of about 450 °C. Mixed gas of $H_2$/Ar with a flow rate of 15/80 sccm was used as the carrier gas. The furnace was ramped to 750 °C at a rate of 50 °C/min and held there for about 5 min to allow the growth of $MoSe_xTe_{2-x}$ crystals. And then, the reaction system was naturally cooled down to room temperature. All reagents were purchased from Alfa Aesar with purity exceeding 99%. By varying the Se and Te weight ratio of the precursor, $MoSe_xTe_{2-x}$ crystals with different Se:Te ratio can be obtained.

**Raman Characterization**. Raman measurements with an excitation laser of 532 nm were performed using a WITEC alpha 300R Confocal Raman system. Before the characterization, the system was calibrated with the Raman peak of Si at 520 $cm^{-1}$. The laser power is less than 1 mW to avoid overheating of the samples.

**STEM Characterization.** The STEM samples were prepared with a poly (methyl methacrylate) (PMMA) assisted method. A layer of PMMA of about 1 μm thick was firstly spin-coated on the wafer with $MoSe_xTe_{2-x}$ samples deposited, and then baked at 180 °C for 3min. The wafer was then immersed in NaOH solution (1M) overnight to etch the $SiO_2$ layer. After lift-off, the PMMA/$MoSe_xTe_{2-x}$ film was transferred into distilled (DI) water for several cycles to rinse off the residual contaminants, and then it was fished by a TEM grid (Quantifoil Au grid). The transferred specimen was dried naturally in ambient environment, and then dropped into acetone overnight to dissolve the PMMA coating layers. The STEM imaging on $MoSe_xTe_{2-x}$ was performed on a FEI Titan Themis with a X-FEG electron gun and a DCOR aberration corrector operating at 60 kV. The inner and outer collection angles for the STEM images ($\beta1$ and $\beta2$) were 54 and 143 mrad, respectively. The convergence semi-angle of the probe is 31 mrad. The beam current was about 20 pA for the ADF imaging. All imaging was performed at room temperature.

**XPS Characterization.** XPS spectra were collected on a PHI Quantera II spectrometer using monochromatic Al-K$\alpha$ (h$\upsilon$ = 1486.6 eV) radiation, and the binding energies were calibrated with C1s binding energy of 284.8 eV. The analysis of peak fitting was performed on XPSPEAK41 software.

**Devices fabrication and transport measurement.** Few-layer Se-doped $MoSe_xTe_{2-x}$ films were firstly identified by their color contrast under optical microscopy. Then small markers were fabricated using standard e-beam lithography (EBL) near the identified sample for subsequent fabrication of Hall-bar devices. To obtain a clean interface between the electrodes and the sample, in situ argon plasma was employed to remove the resist residues before metal evaporation without breaking the vacuum. The Ti/Au (5/70 nm) electrodes were deposited using an electron-beam evaporator followed by lift-off in acetone. Transport experiments were carried out with a standard four-terminal method from room temperature to 0.3 K in a top-loading Helium-3 refrigerator equipped with a 15 T superconducting magnet. A standard low-frequency lock-in technique was used to measure the resistance with an excitation current of 10 nA.

**Data availability.** The data that support the findings of this study are available from the corresponding author upon reasonable request.

**Acknowledgements**

This work has been supported by the National Basic Research Program of China from the MOST under the grant Nos. 2016YFA0300600 and 2015CB921101, by the NSFC under the grant Nos. 11527806 and 11874406. Research in Singapore was funded by the AcRF Tier 3 (2018-T3-1-002), MOE Tier 2 MOE2016-T2-2-153, MOE2015-T2-2-007 and the A*Star QTE Programme. This research is partially supported by the Science, Technology and Innovation Commission of Shenzhen Municipality (No. ZDSYS20170303165926217).

**Author contributions**

P.L., J.C., and J.Z. contributed equally to this work. G.L. conceived and supervised the project; P.L. and J.C. fabricated the devices and carried out the transport measurements; J.Z. synthesized the samples; J.L. did the measurements and data analysis on STEM; G.L. and P.L. prepared the manuscript with input from J.L., J.Z., and Z.L. All the authors discussed the results and commented on the manuscript.

**Additional information**

Supplementary information is available in the online version of the paper. Reprints and permissions information is available online at www.nature.com/reprints.

Correspondence and requests for materials should be addressed to J.Z., J.L., Z.L. or G.L.

**Competing financial interests**

The authors declare no competing financial interests.

**Figure and Figure Caption**

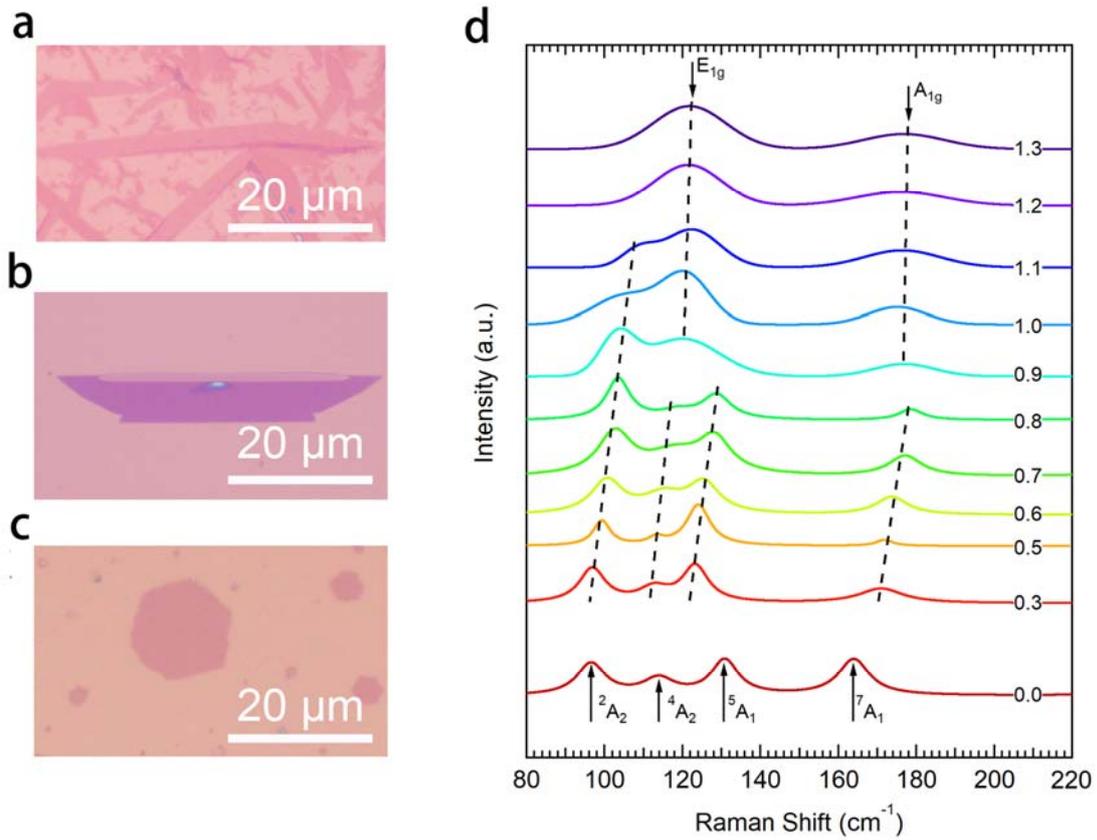

**Figure 1 | Characterizations of Se-substituted MoTe₂ thin films. a**, Optical images of as-synthesized Se-substituted MoTe₂ thin films. The scale bar is 20 μm. **b**, Raman spectra of few-layer MoSe$_x$Te$_{2-x}$ samples with different Se concentrations 0≤ $x$ ≤1.3.

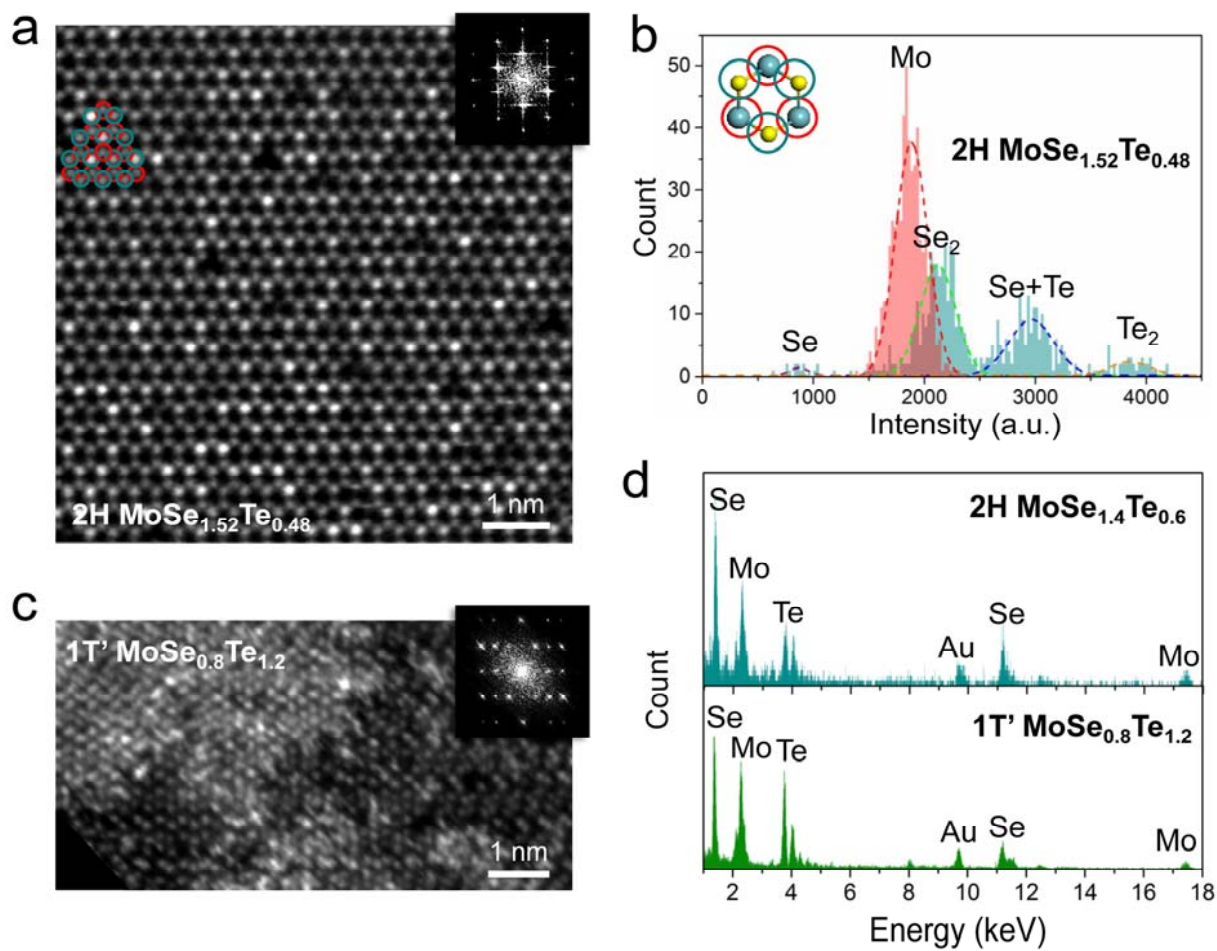

**Figure 2 | High resolution STEM and EDS characterizations of the MoSe$_x$Te$_{2-x}$ monolayers.**
**a**, Atom-resolved STEM image of the CVD-grown MoSe$_x$Te$_{2-x}$ monolayer in 2$H$ phase. The lattice has a hexagonal structure with the brightest spots as Te$_2$ columns. The local chemical stoichiometry is calculated accurately to be MoSe$_{1.52}$Te$_{0.48}$ by mapping the chemical identity of each atomic column through their image intensity. The overlaid red and green circles highlight the cation and anion lattice sites, respectively. The inset is the FFT of the whole image, showing the hexagonal symmetry of the lattice. **b**, Intensity histogram of the mapped atomic columns in (**a**). The cation site has a single peak labelled as Mo while the anion site splits into four peaks, which is assigned (from dark to bright) as Se vacancy, Se$_2$, Se+Te and Te$_2$ columns, respectively.

**c,** Atom-resolved STEM image of CVD-grown MoSe$_x$Te$_{2-x}$ monolayer in 1$T$' phase. The chemical stoichiometry is estimated by EDS as MoSe$_{0.8}$Te$_{1.2}$ with the corresponding FFT image shown as the inset, confirming the quasi-one-dimensional feature of the lattice. **d,** Comparison of the EDS of the MoSe$_x$Te$_{2-x}$ layers in the 2$H$ and the 1$T$' phase, evidencing the phase transition occurs as the Te concentration increase. The Mo K edge is used to normalize the two spectra.

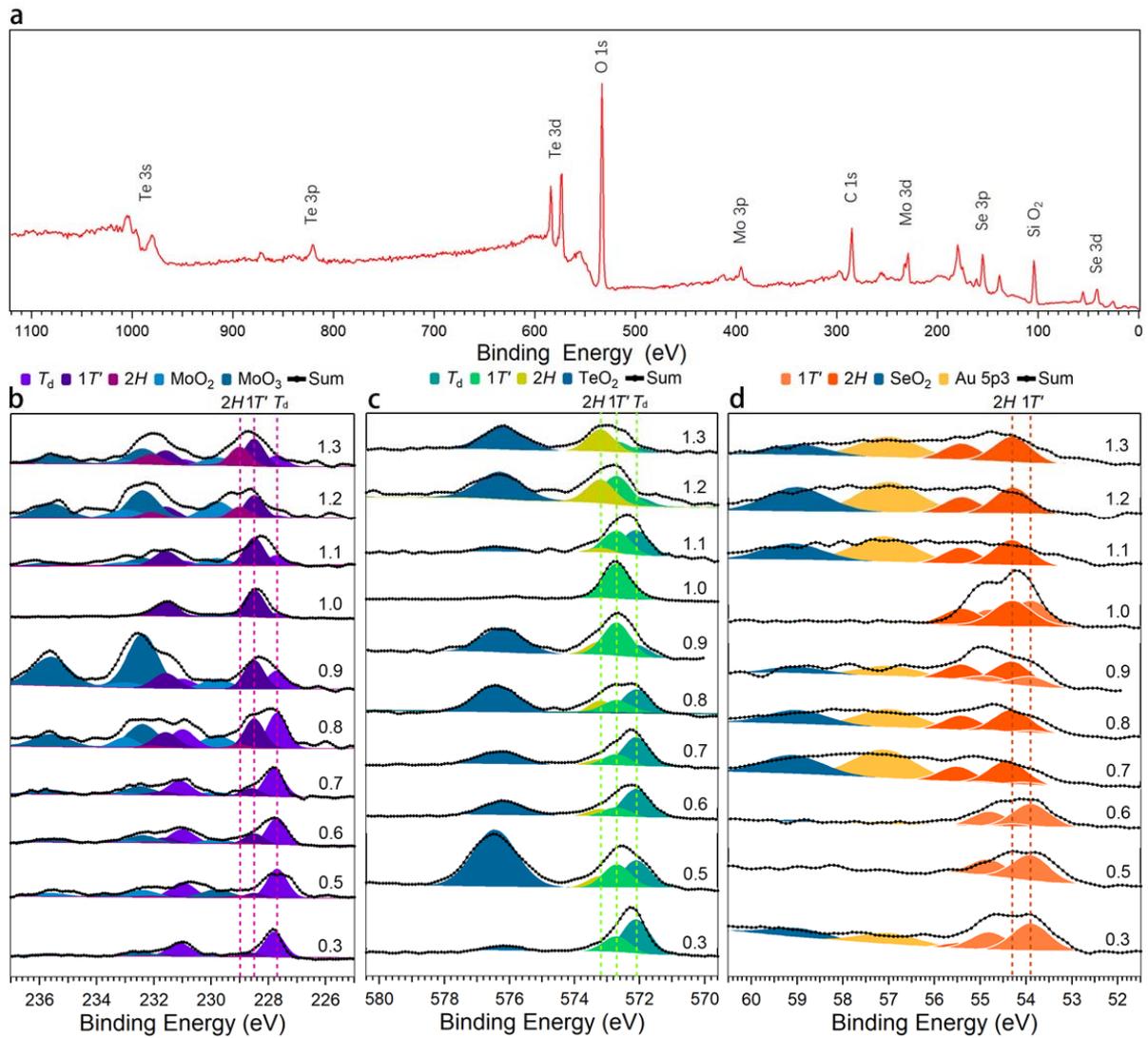

**Figure 3 | XPS characterization of few-layer MoSe$_x$Te$_{2-x}$ with different Se concentrations x.** **a,** A typical XPS spectra of few-layer MoSeTe sample. The peaks of Mo, Se, and Te are from the sample, while the signals of C and O come from the charge reference and the SiO$_2$ substrate, respectively. **b~d**, XPS Spectra of the Mo, Se, and Te 3d binding energies of the few-layer MoSe$_x$Te$_{2-x}$ samples with 0.3≤ *x*≤1.3, respectively. The peaks of MoO$_2$, MoO$_3$, SeO$_2$, and TeO$_2$ are due to the sample oxidation during microfabrication and sample transfer process.

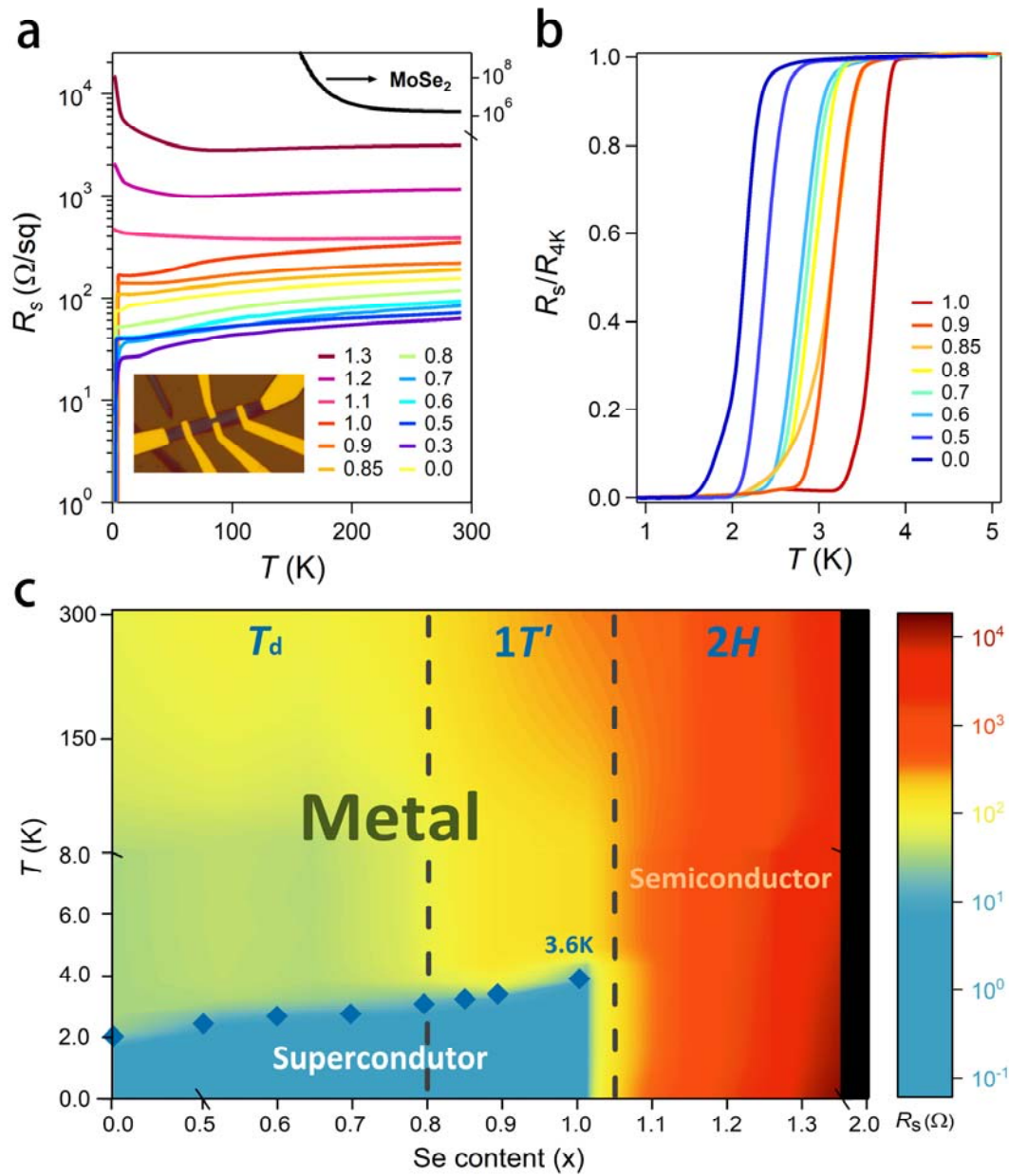

**Figure 4 | Transport characterization of few-layer MoSe$_x$Te$_{2-x}$ thin films with different Se content x. a**, Temperature dependence of the sheet resistance $R_s$ of MoSe$_x$Te$_{2-x}$ samples with different Se contents. Inset: A typical Hall bar device used for low-temperature magnetotransport measurements. **b,** Superconducting transition in few-layer MoSe$_x$Te$_{2-x}$ samples with x⩽1.0. $T_c$ is enhanced from 0.1 K in pristine bulk MoTe$_2$ to 3.6 K in few-layer MoSeTe. **c,** Phase diagram of Se-substituted MoTe$_2$ thin film samples. A phase evolution from the metallic $T_d/1T'$ to the

semiconducting 2*H* phase is observed at a critical substitution content x=1.0. The superconducting transition temperature $T_c$ rises steadily with the Se content to 3.6 K at x=1.0.

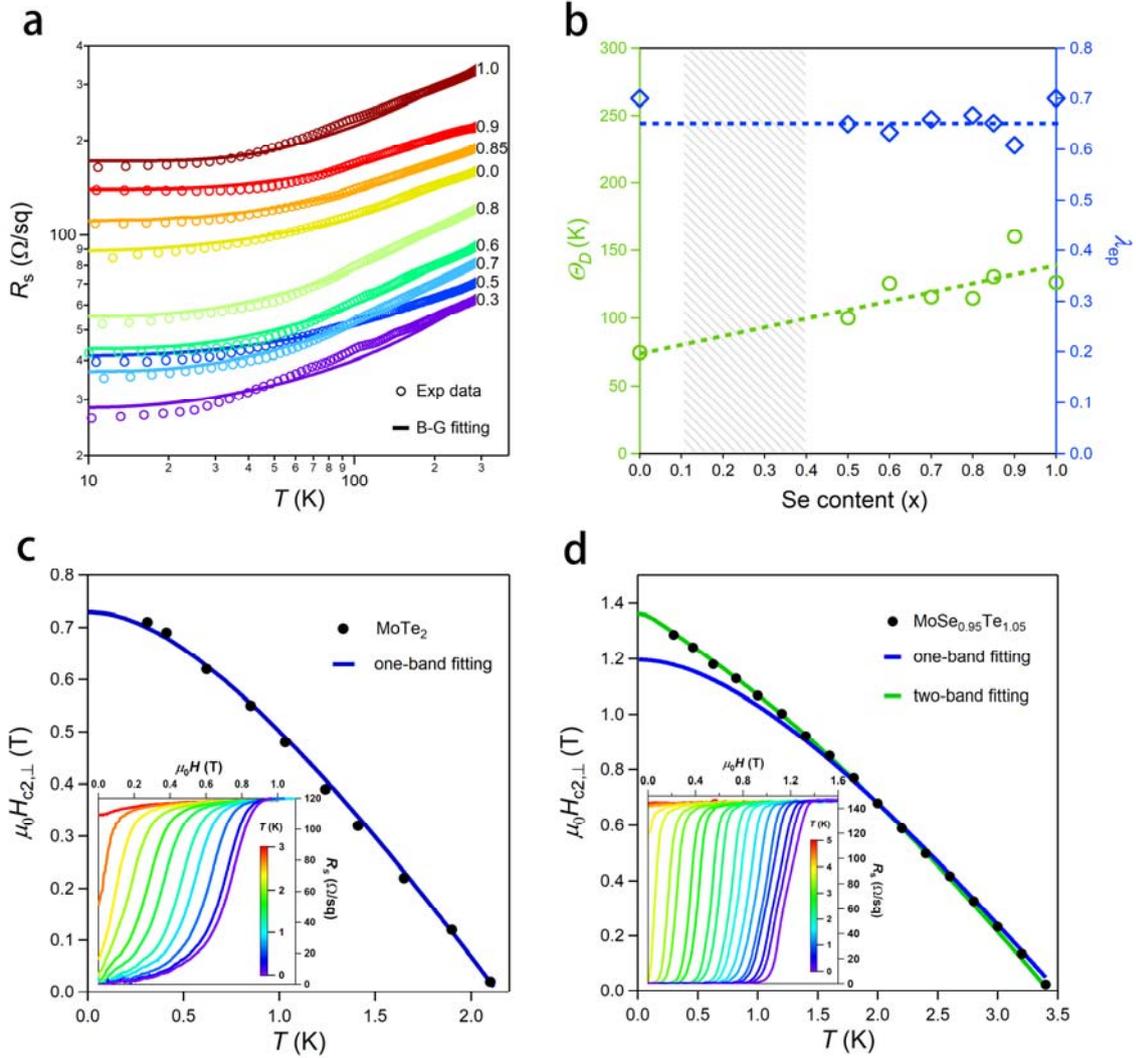

**Figure 5 | Two-band Fitting. a,** Temperature dependence of the normal state sheet resistance $R_s$ of MoSe$_x$Te$_{2-x}$ with x⩽1.0. The solid lines represent the fitting curves with the Bloch-Grüneisen formula. **b,** Se concentration dependence of Debye temperature $\Theta_D$ (left axis) and electron-phonon coupling constant $\lambda_{ep}$ (right axis). **c, d,** Temperature dependence of perpendicular upper critical field $\mu_0 H_{c2,\perp}$ in pristine MoTe$_2$ and MoSe$_{0.95}$Te$_{1.05}$ thin films. The solid blue (green) line represents the fit with the one-band (two-band) model. In the two band model, the coupling constant $\lambda_{11} \approx 0.95$, $\lambda_{22} \approx 0.31$, $\lambda_{12} \approx 0.11$, $\lambda_{21} \approx 0.11$. The inserts of (**c**) and (**d**) are superconducting resistive transition of pristine MoTe$_2$ and MoSe$_{0.95}$Te$_{1.05}$ thin films in perpendicular magnetic fields. The dashed lines indicate the $R_N/2$. $\mu_0 H_{c2}$ is

determined as the intercept between dashed lines and $R_s$ curves. **b.** Temperature dependence of the upper critical field fitting by one-band modal in MoTe$_2$.

**Table I.** Parameters of the measured MoSe$_x$Te$_{2-x}$ samples with different Se content x: room-temperature sheet resistance $R_s$(300K), low-temperature $R_0$, sample thickness $d$, structural phase, the normalized XPS intensities of Mo, Se, and Te.

| x | $R_s$(300K) (Ω) | $R_0$ (Ω) | $d$ (nm) | phase | Mo (I/S) | Se (I/S) | Te (I/S) |
|---|---|---|---|---|---|---|---|
| 0 | 166 | 116 | 8 | $T_d$ | - | - | - |
| 0.3 | 63 | 26 | 60 | $T_d$ | 4580 | 1598 | 7973 |
| 0.5 | 72 | 39 | 20 | $T_d$ | 1611 | 834 | 2374 |
| 0.6 | 92 | 40 | 70 | $T_d$ | 7449 | 4534 | 10336 |
| 0.7 | 85 | 36 | 40 | $T_d$ /1$T'$ | 4281 | 3045 | 5658 |
| 0.8 | 120 | 52 | 15 | 1$T'$ | 3511 | 2851 | 4128 |
| 0.85 | 193 | 110 | 12 | 1$T'$ | 12740 | 10345 | 16003 |
| 0.9 | 221 | 139 | 14 | 1$T'$/2H | 1023 | 959 | 1119 |
| 0.95 | - | 146 | - | 1$T'$/2H | - | - | - |
| 1.0 | 320 | 163 | 40 | 1$T'$/2H | 8699 | 8480 | 8591 |
| 1.1 | 382 | 460 | 5 | 1$T'$/2H | 2134 | 2292 | 1933 |
| 1.2 | 1163 | 2034 | 6 | 2H | 1555 | 1921 | 1248 |
| 1.3 | 3120 | 14844 | 4 | 2H | 2621 | 3382 | 1892 |
| 1.3 | 3120 | 14844 | 4 | 2H | 2621 | 3382 | 1892 |
| 2.0 | 2.6M | - | - | 2H | - | - | - |

*[1]If the sample is a superconductor, $R_0$ is the normal state resistance right above $T_c$, otherwise $R_0$ is the resistance at $T$=1.6 K.

*[2]Mo, Se, Te represent the normalized peak area (I/S) in XPS data for calculating the element stoichiometry. The XPS atomic sensitivity factors (S) of Mo-3d, Se-3d and Te-3d$_{5/2}$ are 3.321, 0.853, and 5.705, respectively.